\documentclass[aps,prl,superscriptaddress,showpacs,twocolumn,10pt]{revtex4-1}

\usepackage{amsmath}
\usepackage{graphicx}
\usepackage{bm}

\usepackage{hyperref}

\begin{document}


\title{Spin dynamics of a confined electron interacting with magnetic or nuclear spins: \\
A semiclassical approach}



\author{Tomasz Dietl}
\email[]{dietl@ifpan.edu.pl}
\affiliation{Institute of Physics, Polish Academy of Sciences,
aleja Lotnik\'{o}w 32/46, PL-02-668 Warszawa, Poland}
\affiliation{Institute of Theoretical Physics, Faculty of Physics, University of Warsaw,
ulica Pasteura 5, PL-02-093 Warszawa, Poland}
\affiliation{WPI-Advanced Institute for Materials Research (WPI-AIMR),
Tohoku University, 2-1-1 Katahira, Aoba-ku, Sendai 980-8577, Japan}

\date{\today}

\begin{abstract}
 %
A physically transparent and mathematically simple semiclassical model is employed to examine dynamics in the central-spin problem.  The results reproduce a number of previous findings obtained by various quantum approaches and, at the same time, provide information on the electron spin dynamics and Berry's phase effects over a wider range of experimentally relevant parameters than available previously. This development is relevant to dynamics of bound magnetic polarons and spin dephasing of an electron trapped by an impurity or a quantum dot, and coupled by a contact interaction to neighboring localized magnetic impurities or nuclear spins. Furthermore, it substantiates the applicability of semiclassical models to simulate dynamic properties of spintronic nanostructures with a mesoscopic number of spins.
\end{abstract}

\pacs{03.65.Yz,73.21.La,75.50.Pp,76.60.Es}

\maketitle

A persistent progress in controlling of ever smaller spin ensembles has stimulated the development of various quantum approaches \cite{Barnes:2012_PRL,Barnes:2011_PRB,Erbe:2010_PRL,Coish:2010_PRB,Cywinski:2009_PRB,Deng:2008_PRB,Bortz:2007_PRB,Zhang:2006_PRB,Yao:2006_PRB,Khaetskii:2003_PRB} and numerical diagonalization procedures \cite{Faribault:2013_PRL,Cywinski:2010_PRB,Al-Hassanieh:2006_PRL,Dobrovitski:2003_PRE}  to the central-spin problem \cite{Gaudin:1976_JdP,Kasuya:1968_RMP}, designed to describe dephasing of a single electron coupled to a spin bath residing within electron's confinement region. These works have been put forward to understand effects of nuclear magnetic moments on electron spin qubits in quantum dots \cite{Bluhm:2011_NP,Barthel:2009_PRL,Koppens:2008_PRL,Petta:2005_S} but they appear also relevant to studies of spin dynamics of a confined electron in dilute magnetic semiconductors (DMSs) at times shorter than intrinsic transverse relaxation time $T_{2}$ of the central and bath spins. However, due to inherent complexity, the quantum models are so far valid in a restricted range of experimental parameters.



In this paper, we reexamine spin dephasing of a confined electron employing a previous semiclassical approach to the central-spin problem \cite{Dietl:1983_PRB,Dietl:1982_PRL}. The proposed model appears similar to more recent semiclassical treatments of electron spin dynamics in the presence of a nuclear spin bath \cite{Hung:2013_PRB,Neder:2011_PRB,Merkulov:2010_PRB,Erlingsson:2004_PRB,Semenov:2003_PRB,Merkulov:2002_PRB}, but its significantly more generic formulation put forward here allows us to consider less restricted ranges of times $t$, magnetic fields $B$, polarizations $p_I$ and lengths $I$ of the bath spins as well as to take into account Berry's phase, polaronic effects, and spin-spin interactions within the bath. These interactions are encoded in the dynamic longitudinal and transverse magnetic susceptibilities $\chi_{\vec{q}}(\omega)$ of the system in the absence of the electron, which---at least in principle---are available experimentally, and constitute the input parameters to the theory. In this way we provide a formalism suitable to describe experimental results in the hitherto unavailable parameter space, allowing also to benchmark various implementations of quantum theory and to establish limitations of the present semiclassical model.

The starting point \cite{Dietl:1983_PRB,Dietl:1983_JMMM} is the electron spin Hamiltonian $\hat{\vec{s}}\vec{\Delta}$ with eigenvalues describing spin-split electron energies $\pm \frac{1}{2}\Delta$, where in the presence of a collinear magnetic field $\vec{B}$ and an average magnetization $\vec{M}_0$ of bath spins each with a magnetic moment $\mu_I$,
\begin{equation}
\vec{\Delta} =  g^*\mu_B\vec{B} + \frac{JI}{|\mu_I|}(\vec{M}_0 + \sum_{\vec{q}}b_{\vec{q}}\vec{\eta}_{\vec{q}}).
\label{eq:Delta}
\end{equation}
Here $g^*$ is the electron effective Land\'e factor; $J$ is either the $s$--$d$ exchange integral $\alpha$ or, in terms of the hyperfine interaction energy ${\cal{A}}$, $J = {\cal{A}}/N_0$, where $N_0$ is the cation concentration (inverse unit cell volume in zinc-blende semiconductors); $b_{\vec{q}}$ and $\vec{\eta}_{\vec{q}}$ are Fourier components of the square of the electron envelope function $|\psi(\vec{r})|^2$ and of bath magnetization fluctuations $\vec{M}(\vec{r}) - \vec{M}_0$, respectively. If the electron localization length is much longer than an average distance between the bath spins the summation over $q$ can be extended to infinity but otherwise an appropriate cutoff $q_{\text{max}}$ should be implemented \cite{Jaroszynski:1985_SSC}.

Except for the immediate vicinity to spin ordering temperature, the distribution of $\vec{\eta}_{\vec{q}}$ is, to a good approximation, gaussian for any mixed state that can be described by spin temperature with the variance, according to the fluctuation-dissipation theorem, determined by an appropriate integral over $\omega$ of the imaginary part of $\chi_{\vec{q}}(\omega)$. Since $\vec{\Delta}$ is linear in $\vec{\eta}_{\vec{q}}$, the central limit theorem implies that the distribution of $\vec{\Delta}$ in the absence of the electron, $P_I(\vec{\Delta})$, is also gaussian,
\begin{equation}
P_I(\vec{\Delta}) = Z_I^{-1}\exp[-(\Delta_z - \Delta_0)^2/2\sigma_{\|}^2 - \vec{\Delta}_{\bot}^2/2\sigma_{\bot}^2],
\label{eq:PI}
\end{equation}
where $Z_I = (2\pi)^{3/2}\sigma_{\|}\sigma_{\bot}^2$ is the probability normalizing constant insuring that $\int d\vec{\Delta} P_I(\vec{\Delta}) = 1$ and $\Delta^2 = \Delta_z^2 + \vec{\Delta}_{\bot}^2$ with the $z$-axis taken along the magnetic field.
The three parameters characterizing
cubic systems are then given by \cite{Dietl:1983_PRB,Dietl:1983_JMMM}
\begin{eqnarray}
\Delta_0 &=& g^*\mu_BB + \frac{JI}{|\mu_I|}M_0,
\label{eq:Delta0} \\
\sigma^2_{\|(\bot)} &=& \left(\frac{JI}{\mu_I}\right)^2k_{B}T\sum_{\vec{q}}\tilde{\chi}_{\|(\bot)\vec{q}}|b_{\vec{q}}|^2/\mu_0.
\label{eq:sigma}
\end{eqnarray}
If energies of bath magnetic excitations are smaller than the thermal energy $k_{B}T$ and their $q$-dependence irrelevant then
$\tilde{\chi}_{\|} = \mu_0\partial M_0/\partial B$ and $\tilde{\chi}_{\bot} = \mu_0M_0/B$, i.e., except for the region near magnetization saturation, $\sigma_{\|} \simeq \sigma_{\bot}$.

By adding the free energy of the electron,
so that
\begin{equation}
P(\vec{\Delta}) = Z^{-1}2\cosh(\Delta/2k_BT)P_I(\vec{\Delta}),
\label{eq:P}
\end{equation}
the bound magnetic polaron (BMP) effects in thermal equilibrium are taken into account. Since non-scalar spin-spin interactions break spin momentum conservation, the BMP formation time is of the order of $T_{2(I)}$ of the bath spins. Hence, up to a time scale of $T_{1(I)}$, adiabatic rather than isothermal magnetic susceptibilities describe energetics of the system \cite{Dietl:1995_PRL}.

For a typical number of bath spins within the confinement region, $N_I \simeq 50$,
this model was found entirely equivalent to quantum approaches developed independently \cite{Ryabchenko:1983_JETP,Heiman:1983_PRB}. Moreover, the probability distribution of spin splitting $P(\Delta)$, obtained from $P(\vec{\Delta})$ by integration over polar and azimuthal angles \cite{Dietl:1983_PRB,Dietl:1982_PRL}, together  with appropriate optical selection rules \cite{Dietl:1983_JMMM}, describe with no adjustable parameters the position, width, and shape of spin-flip Raman scattering lines for electrons bound to shallow donors in various II-VI DMS \cite{Peterson:1985_PRB,Alov:1983_JETP,Dietl:1982_PRL,Alov:1981_JETPL,Nawrocki:1981_PRL}. Furthermore, the contribution to the system free energy brought by the electron \cite{Dietl:1983_PRB,Dietl:1982_PRL}, $F = -k_BT\ln Z$,  provides magnetic susceptibilities of BMPs \cite{Dietl:1983_PRB,Dietl:1982_PRL}, whose magnitudes are in quantitative agreement with experimental values as a function of temperature \cite{Dietl:1982_PRL} and the donor electron concentration \cite{Wojtowicz:1993_PRL}.

While in DMSs a full account of the spin-spin interactions is essential [and implemented by using experimental values of $M_0(B,T)$ in Eqs.\,\ref{eq:Delta0} and \ref{eq:sigma}], in the case of a nuclear bath interactions among spins are often negligible. In such a case, in terms of polarization of spins $p_I$, i.e., the paramagnetic Brillouin function B$_I$,
\begin{eqnarray}
\Delta_0 &=& g^*\mu_BB + x_IN_0JIp_I,
\label{eq:Delta0p} \\
\sigma^2_{\|(\bot)} &=& (x_IN_0J)^2 I(I+1) f_{\|(\bot)}/3N_I,
\label{eq:sigmap}
\end{eqnarray}
where $x_I$ is the fractional content of the spin $I$ in the crystal lattice (for instance, in GaAs, $I = 3/2$ and $x_I = 2$ if an average value
of the hyperfine coupling energy ${\cal{A}} = N_0J$ is used; see, e.g., Ref.~\,\onlinecite{Hung:2013_PRB}),
\begin{equation}
N_I = x_IN_0/\sum_q |b_{\vec{q}}|^2 = x_IN_0/\int d\vec{r}|\psi(\vec{r})|^4,
\label{eq:N}
\end{equation}
$f_{\|} = [3/(I+1)]\partial p_I/\partial y$ and $f_{\bot} = [3/(I+1)]p_I/y$, where $y$ is determined by the implicit relation,
 \begin{equation}
 p_I = [(2I + 1)/2I]\coth[(2I + 1)y/2] - (1/2I)\coth(y/2).
 \label{eq:pI}
\end{equation}
As seen, $f_{\|(\bot)} \rightarrow 1$ for $p_I \rightarrow 0$ and $f_{\|(\bot)} \rightarrow 0$ for $p_I \rightarrow 1$. The last
describes narrowing of the spin distribution when polarization is enhanced by, e.g., a magnetic field or, dynamically, by the Overhauser effect \cite{Bluhm:2010_PRL,Foletti:2009_NP}. We also note that in the case of several uncoupled spin species $\alpha$ (e.g., a heteronuclear system, the case of GaAs) $P_I(\Delta) = \prod_{\alpha}P_{I_{\alpha}}(\Delta_{\alpha})$.

In the absence of polaronic effects, the most probable value of spin splitting $\bar{\Delta}$ corresponds to a maximum of $P_I(\Delta)$. Hence, in terms of $\Delta_0$ and for $\sigma_{\|} = \sigma_{\bot} =\sigma$, $\bar{\Delta}$ is determined by \cite{Dietl:1982_PRL,Dietl:1983_PRB}
\begin{equation}
\bar{\Delta}^2 -\bar{\Delta}\Delta_0\coth(\bar{\Delta}\Delta_0/\sigma^2) - \sigma^2 = 0.
\label{eq:Dp}
\end{equation}
This expression, by describing corrections to spin splitting $\Delta_0$ brought about by magnetization fluctuations, generalizes for an arbitrary value of $\Delta_0$ the formula obtained by quantum methods  in the regime $\Delta_0 \gg \sigma$ (and referred to as a Lamb shift) \cite{Barnes:2012_PRL,Barnes:2011_PRB,Coish:2010_PRB,Deng:2008_PRB,Bortz:2007_PRB}. In this range, according to Eq.\,\ref{eq:Dp}, $\bar{\Delta}= \Delta_0 + \sigma^2/\Delta_0$. If only transverse magnetization fluctuations are taken into account, $\bar{\Delta}_{\text{n}}= \Delta_0 + \sigma_{\bot}^2/2\Delta_0$, the result in agreement with the corresponding determination of the Lamb shift for the narrowed spin bath \cite{Coish:2010_PRB}.

Having specified the distribution of bath spin orientations, we move to electron dephasing that is characterized by the time evolution of the off-diagonal component of the $s = \frac{1}{2}$ density matrix $\hat{\rho}(t)$ normalized by its initial value, $w(t) = \rho_{12}(t)/\rho_{12}(0)$. We describe this dynamics (free induction decay - FID) by quantum Liouville's equation with the Hamiltonian $H_s =\hat{\vec{s}}\vec{\Delta}(t)$, where $\vec{\Delta}(t)$ is a classical field. Its time dependence is determined by: (i) the presence of the electron, the effect we neglect (see Ref.\,\onlinecite{Merkulov:2010_PRB}) and (ii) internal bath dynamics. According to Eq.\,\ref{eq:Delta} and the fluctuation-dissipation theorem relevant information is provided by correlation functions $\langle \Delta_i(t)\Delta_j(t')\rangle$ determined, in turn,  by the Fourier transform of $\chi^{\prime \prime}_{ij\vec{q}}(\omega)$ available from paramagnetic or nuclear resonance studies in the case of DMSs and non-magnetic systems, respectively.

One of models that have been put forward in dephasing studies \cite{Hung:2013_PRB,Neder:2011_PRB,Cywinski:2009_PRB,Yao:2006_PRB} consists of assuming $H_s$ in a truncated form, $H_s=\hat{s}_z[\Delta_z(t) + \Delta^2_{\bot}(t)/2\Delta_0]$, which may be justified in the region $\Delta_0 \gg \sigma_{\|(\bot)}$, where a relative contribution of $\Delta_{\bot}(t)$ to the most probable $\vec{\Delta}$ is relatively small.
It is straightforward to reproduce within our formalism theoretical results obtained by using such an effective Hamiltonian \cite{Hung:2013_PRB,Neder:2011_PRB,Liu:2010_AP}.

\begin{figure}[tb]
   \hbox to \hsize{%
    \hfill\includegraphics[scale=0.165]{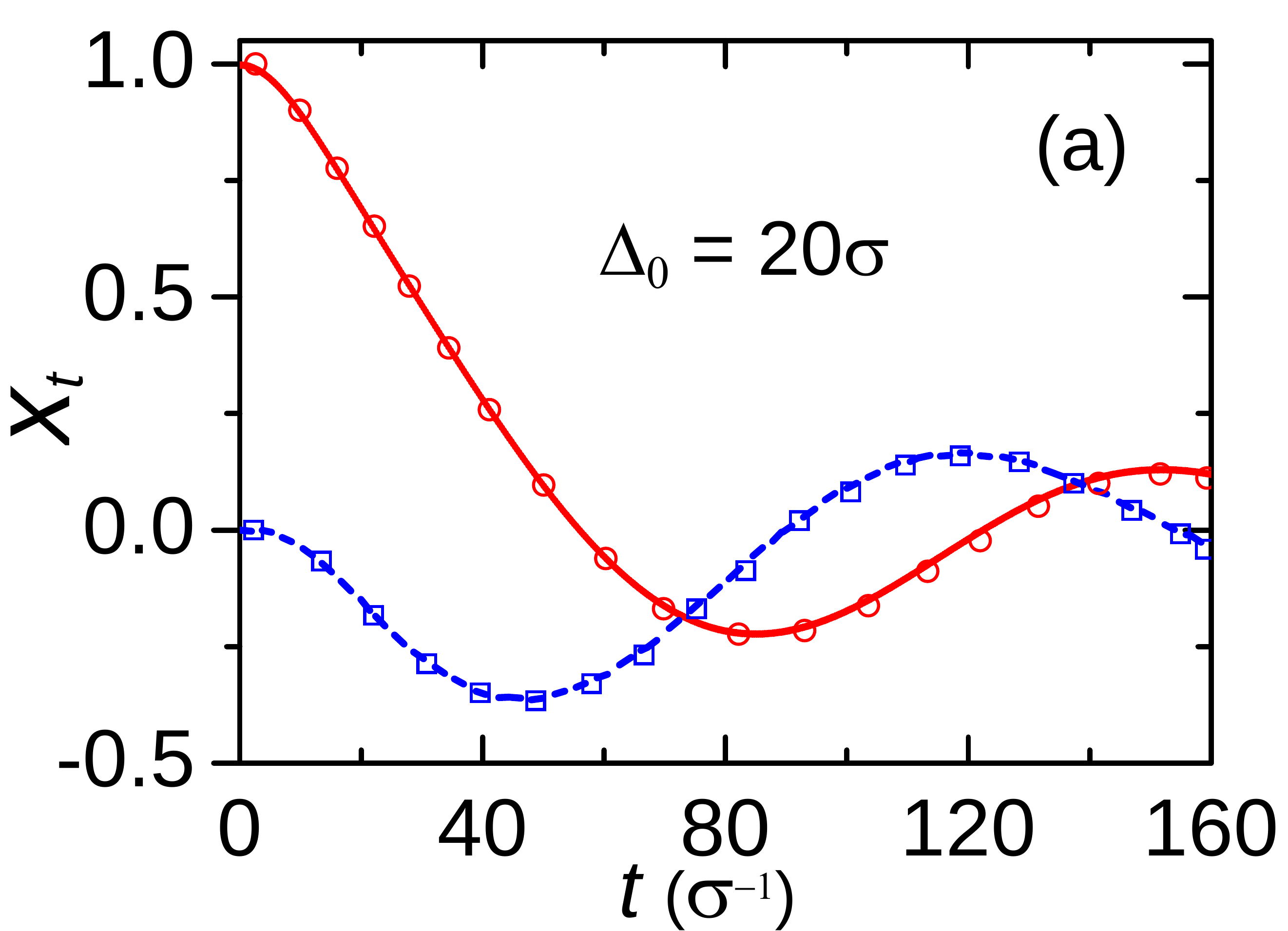}\hfill
    \hfill\includegraphics[scale=0.165]{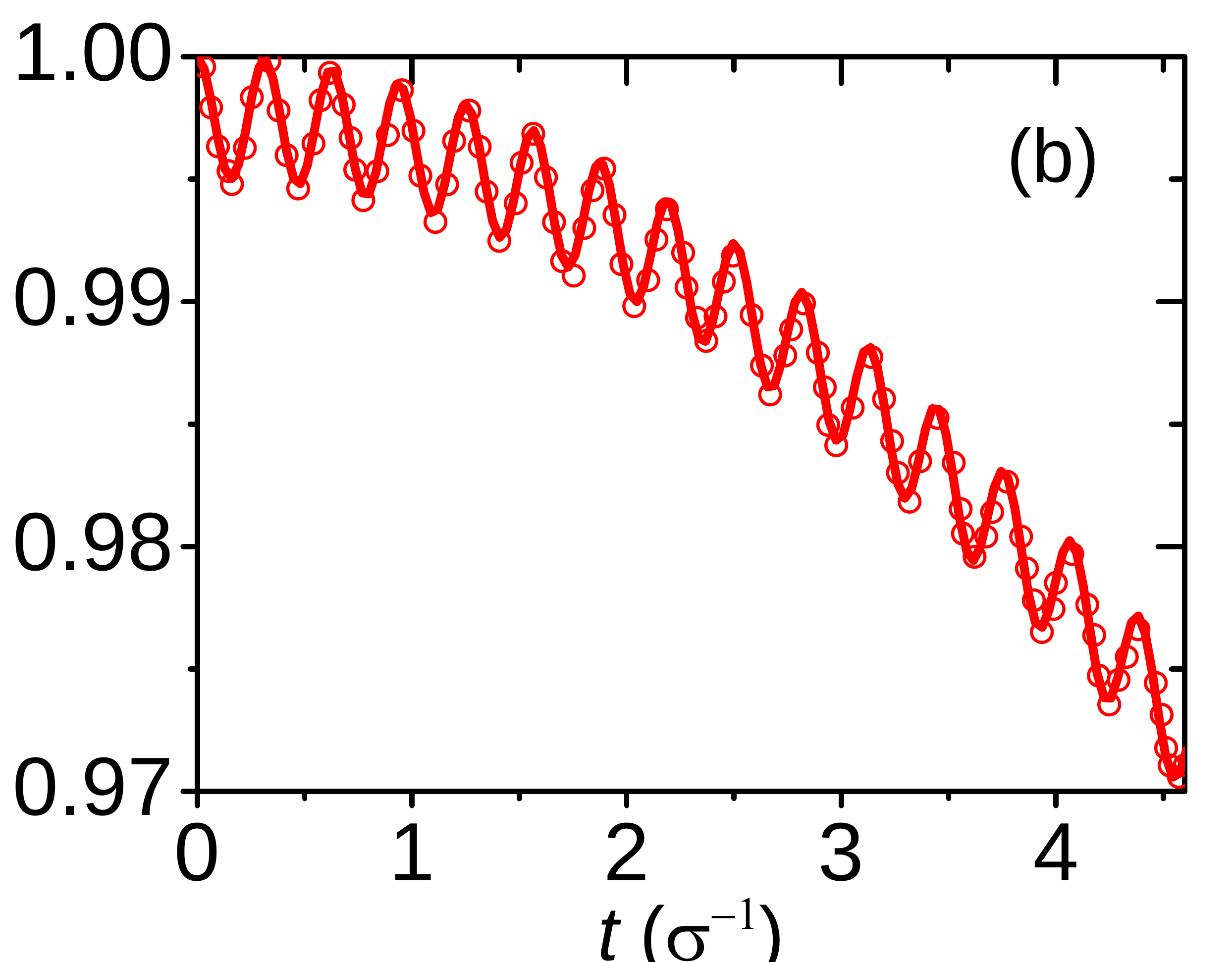}\hfill}
 	\caption{(a) Time dependence of the coherence function in the rotating frame and in the narrowed bath case, $x_t = \exp(-it\bar{\Delta})w_t$, where $w_t$ is given in Eq.\,\ref{eq:w3} for $\Delta_0 = 20\sigma_{\bot}$ and $\omega_I = 0$. Real and imaginary parts of $x_t$ are shown by solid and dashed lines, respectively.  (b) Real part in an expanded scale. Points denote data displayed as lines in Figs.\,1 and 2 in Ref.\,\onlinecite{Barnes:2011_PRB} and referred to as "exact".}
  \label{fig:x}
\end{figure}

We turn, therefore, to a theory appropriate for an arbitrary ratio $\Delta_0/\sigma_{\|(\bot)}$. We consider a homospin system in the regime
$t < T_{2(I)}$ and $t < N_I/|x_IN_0J|$. In this range, the time dependence of $\vec{\Delta}$ is neither affected by interactions among bath spins nor by the molecular field of the central spin. It comes solely from the field-induced precession (with the frequency $\omega_I = -\mu_IB/I$) that we treat in the adiabatic approximation \cite{Romero:2002_PA,Erlingsson:2004_PRB}. For experiments involving sequential measurements of FID for a single dot or donor impurity we are interested in $w(t)$ averaged over possible initial values of $\vec{\Delta}$. After integrating over the azimuthal angle $\varphi$ we arrive to the coherence function $w_t$ in the form,
\begin{equation}
w_t = 2\pi\int_{-\infty}^{\infty}d \Delta_z \int_0^{\infty}d \Delta_{\bot} \Delta_{\bot} P_I(\vec{\Delta})(\cos d_t + ix\sin d_t)^2.
\label{eq:w}
\end{equation}
Here, $x \equiv \cos\theta = \Delta_z/\Delta$ and, implementing henceforth $\hbar = 1$,  $d_t = [\Delta + \mbox{sgn}(x)\omega_I(1-|x|)]t/2$, where the term with $\omega_I$ describes geometrical Berry's phase. If the measurement sequence allows for the BMP formation and the BMP energy \cite{Dietl:1982_PRL} $\epsilon_p = \sigma_{\|}^2/4k_BT  > k_BT$ then  $P(\vec{\Delta})$ given in Eq.\,\ref{eq:P} [rather than $P_I(\vec{\Delta})$, Eq.\,\ref{eq:PI}] should be employed in Eq.\,\ref{eq:w}.

If one neglects both the Berry phase effects and the difference between  $\sigma_{\|}$ and  $\sigma_{\bot}$, i.e., if $p_I \lesssim 0.2$, Eq.\,\ref{eq:w} assumes an analytic form for the distribution of $\vec{\Delta}$ described by either $P_I(\vec{\Delta})$ (Eq.\,\ref{eq:PI}) or $P(\vec{\Delta})$ (Eq.\,\ref{eq:P}). In the former case we obtain,
\begin{eqnarray}
w_t &=& 1/\Delta_0^2  + (1 + it/\Delta_0 - 1/\Delta_0^2)\exp(-t^2/2 +it\Delta_0) \nonumber \\
&+& \frac{\sqrt{2\pi}i}{4\Delta_0^3}[\mbox{erf}(t/\sqrt{2}-i\Delta_0/\sqrt{2})
-\mbox{erf}(t/\sqrt{2} +i\Delta_0/\sqrt{2}) \nonumber\\
 &+& 2\mbox{erf}(i\Delta_0/\sqrt{2}]\exp(-\Delta_0^2/2),
\label{eq:w2}
\end{eqnarray}
where $\Delta_0$ and $t$ are in the units of $\sigma$ and $\sigma^{-1}$, respectively. It can be easily shown that  the above equation is not only equivalent to the expression derived previously within the semiclassical approach for $\Delta_0 \rightarrow 0$ \cite{Merkulov:2002_PRB} but is also related to a quantum formula given in Eq.\,3 of Ref.\,\onlinecite{Zhang:2006_PRB}.

\begin{figure}[tb]
   \includegraphics[width=1.0\columnwidth]{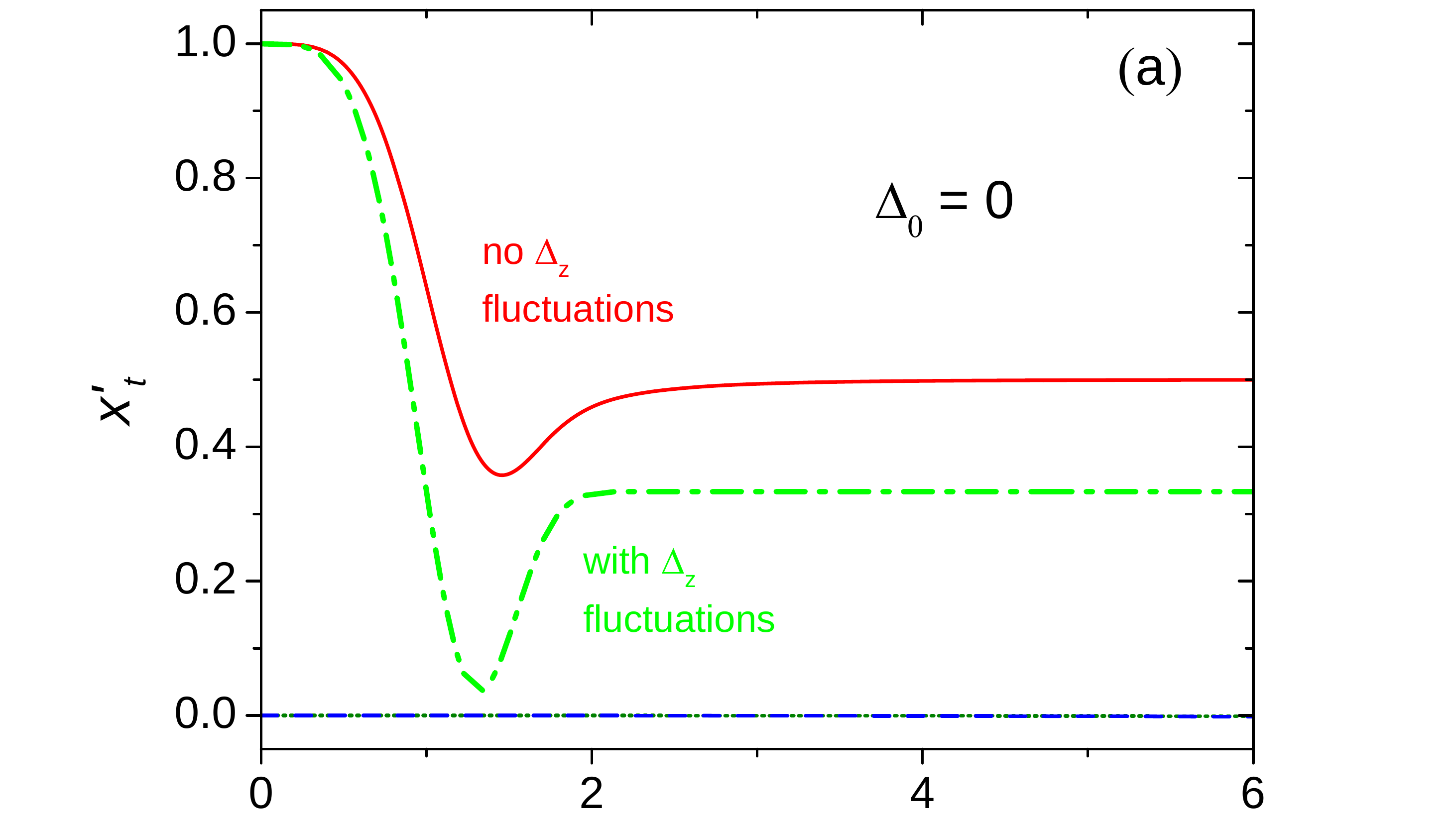}
    \includegraphics[width=1.0\columnwidth]{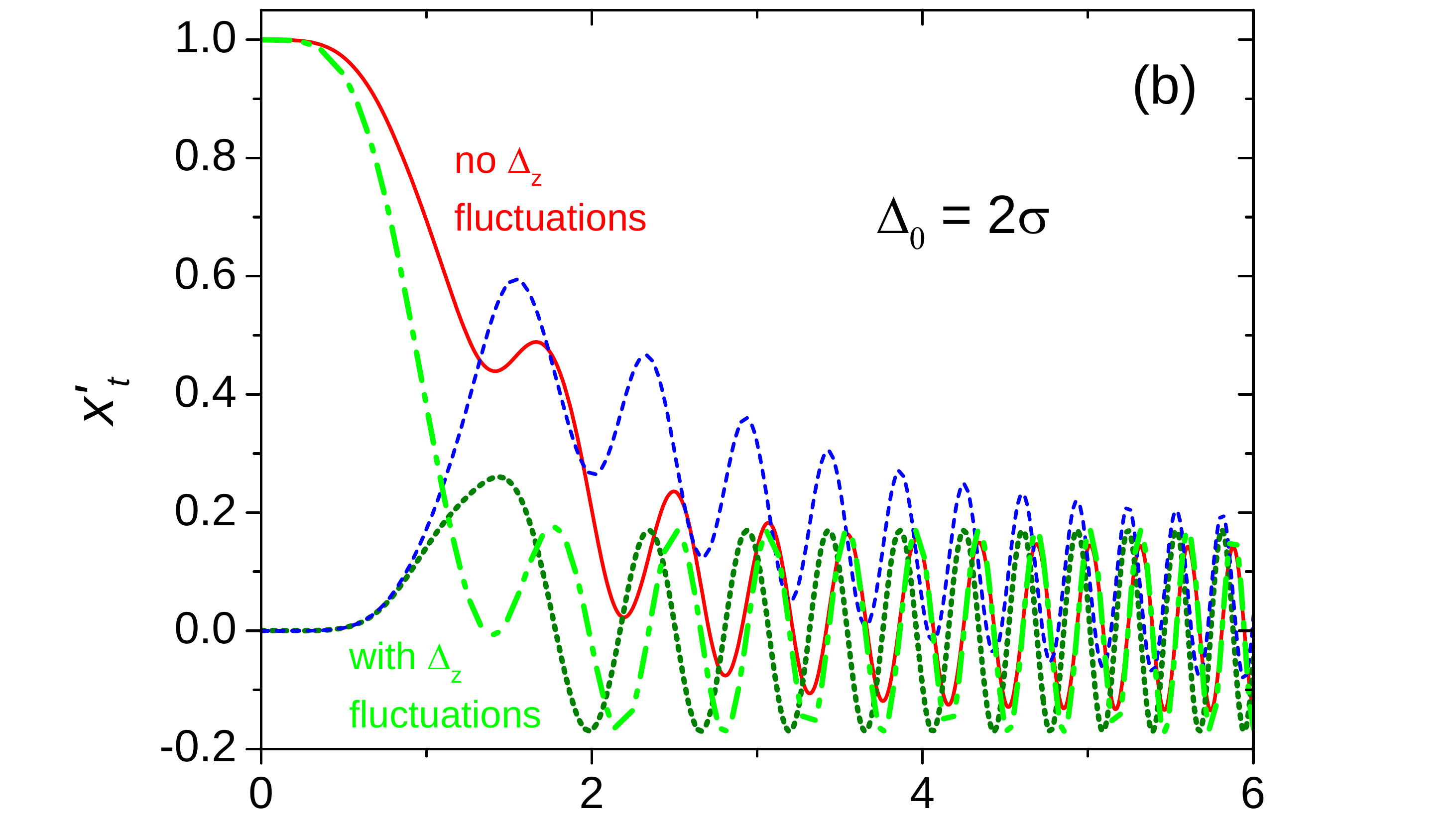}
    \includegraphics[width=1.0\columnwidth]{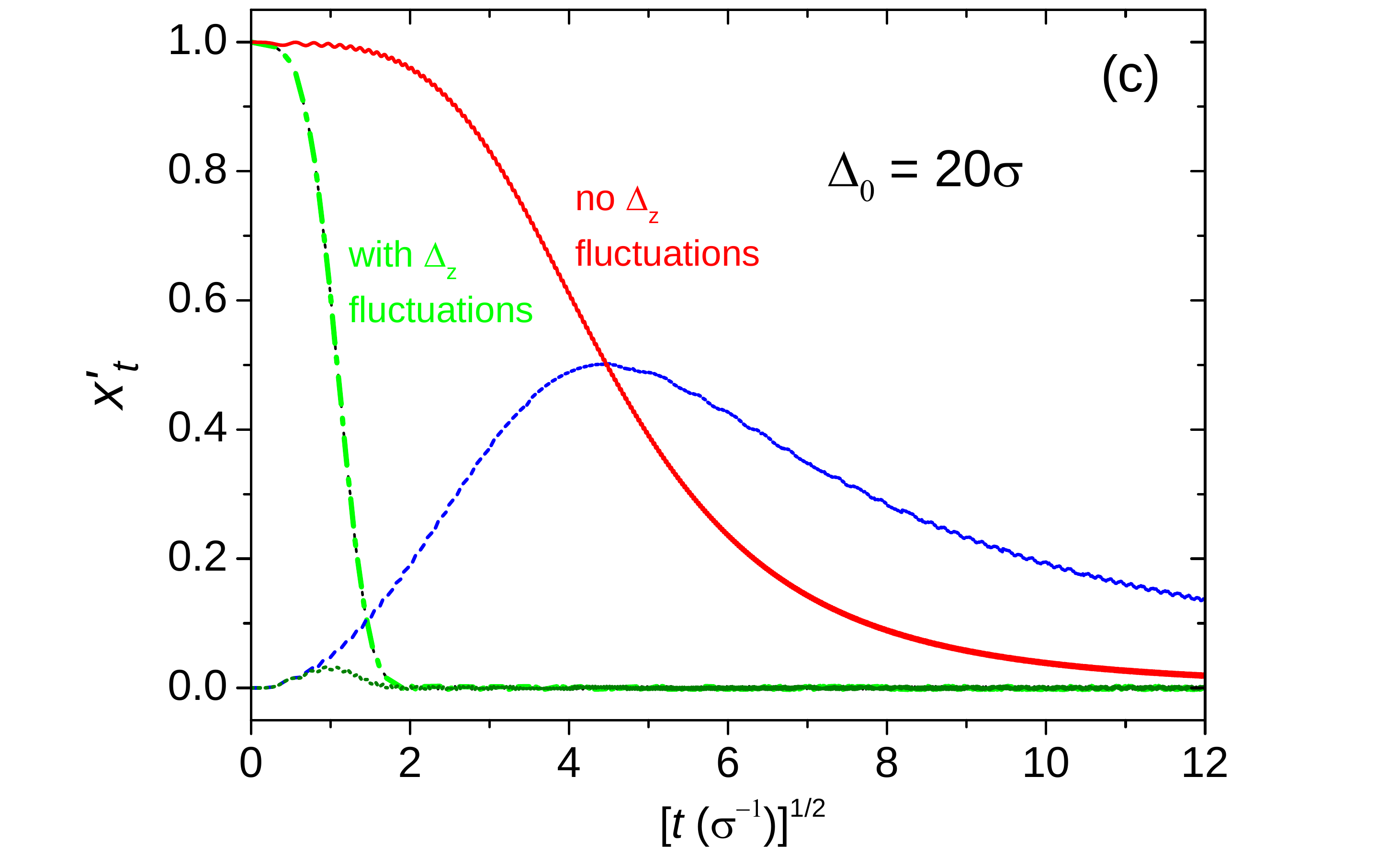}
\caption{Real and imaginary parts of coherence function in the rotating frame, $x^{\prime}_t = \exp(-it\Delta_0)w_t$ as a function of the square root of time in the absence of $\Delta_z$ fluctuations -- narrowed spin bath (Eq.\,\ref{eq:w3}, $\omega_I = 0$; solid and dashed lines) and with $\Delta_z$ fluctuations included (Eq.\,\ref{eq:w2}; dashed-dot and dot lines). Results for $\Delta_0 =  0$, $2\sigma$, and $20\sigma$ are shown in panels (a), (b), and (c), respectively.}
  \label{fig:xp}
\end{figure}

Another relevant case corresponds to the narrowed spin bath, $\sigma_{\|} \rightarrow 0$ and $\sigma = \sigma_{\bot}$. Adopting again units in which $\sigma_{\bot} = 1$, Eq.\,\ref{eq:w} leads to
\begin{equation}
w_t = \int_0^{\infty}d \Delta_{\bot}(\cos d_t + ix\sin d_t)^2\Delta_{\bot}\exp(-\Delta_{\bot}^2/2),
\label{eq:w3}
\end{equation}
where $\Delta_z = \Delta_0$ in $d_t$ and $x$.

We first present results obtained within our model in the parameter region for which quantum theories have been developed, $\Delta_0 \gg \sigma_{\bot}$, $\omega_I = 0$ (no Berry phase effects), and for the case of a narrowed spin bath. The data computed from Eq.\,\ref{eq:w3} for $\Delta_0 = 20\sigma$ are collected in Fig.~\,\ref{fig:x}. Their comparison to findings presented in Figs.\,1 and 2 of Ref.\,\cite{Barnes:2011_PRB}, obtained for $p_I =0$ demonstrates a quantitative equivalence of the semiclassical and quantum approaches in this case. This agreement holds up to $t \gg \sigma^{-1}$.

Encouraged by this finding, we illustrate in Fig.\,\ref{fig:xp} the theoretically expected behavior of the coherence function in the rotating frame, $x^{\prime}_t = \exp(-it\Delta_0)w_t$, for three distinct cases, $\Delta_0 = 0$, $2\sigma$, and $20\sigma$, considering the situation when both transverse and longitudinal magnetization fluctuations are present (Eq.\,\ref{eq:w2}) and also the case of a narrowed spin bath (Eq.\,\ref{eq:w3}), where $\sigma$ corresponds to $\sigma_{\bot}$. A saturation of $x^{\prime}_t$ at 1/3 and at 1/2 for $t \rightarrow \infty$, visible in Fig.\,\ref{fig:xp}(a), was previously noted (Refs.\,\onlinecite{Khaetskii:2003_PRB,Merkulov:2002_PRB} and Ref.\,\onlinecite{Faribault:2013_PRL}, respectively). Oscillations in $x^{\prime}_t$ appearing in Fig.\,\ref{fig:xp}(b,c) have the frequency of $\Delta_0$.

\begin{figure}[tb]
   \includegraphics[width=1.0\columnwidth]{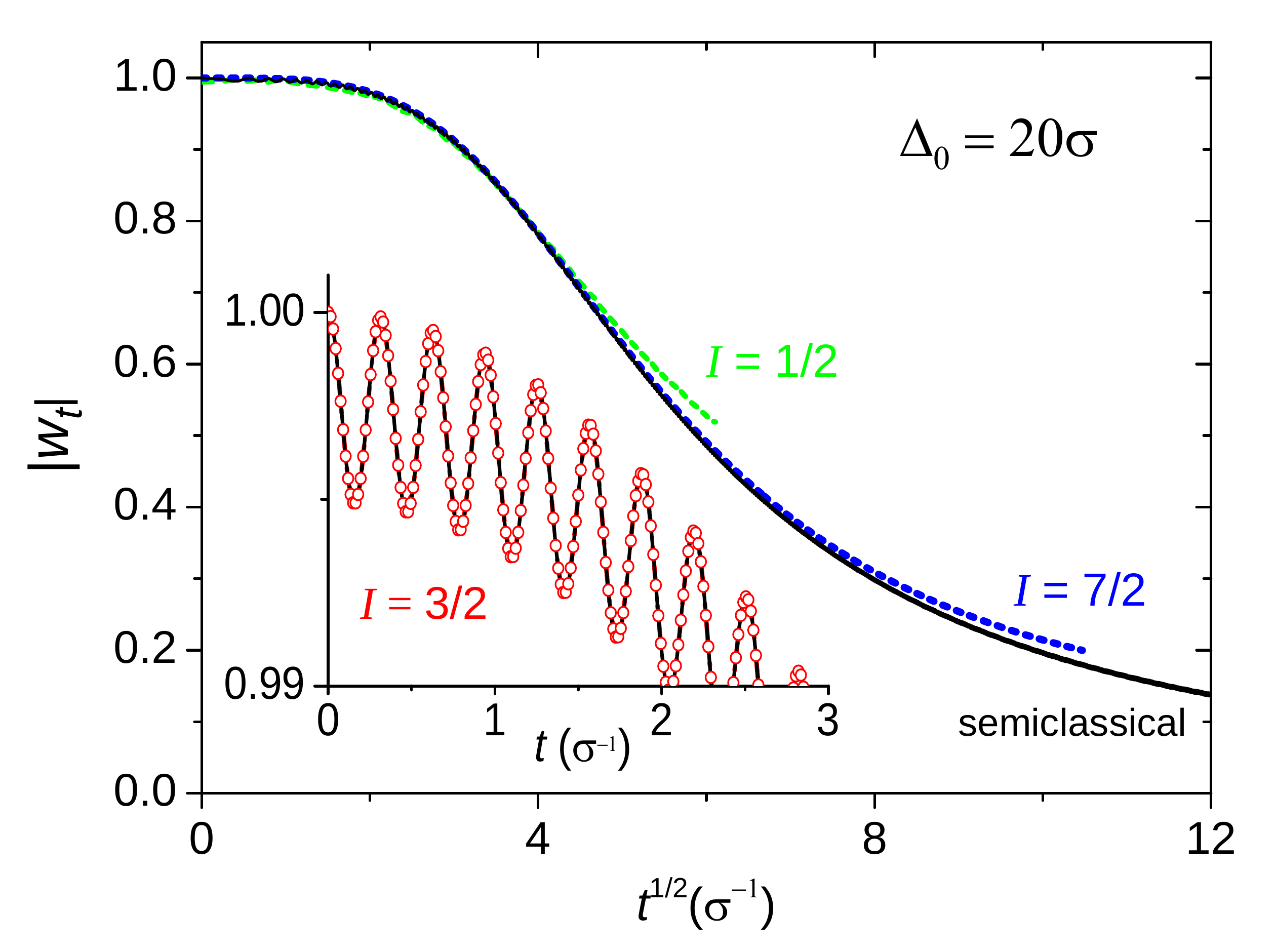}
\caption{Absolute value of coherence function $|w_t|$ for narrowed bath computed within the semiclassical theory from Eq.\,\ref{eq:w3} (valid for arbitrary bath polarization $p_I$ and spin $I$) assuming $\Delta_0 =  20\sigma_{\bot}$ and $\omega_I = 0$  as a function of the square root of time (solid line). Predictions of the quantum model (Eq.\,\ref{eq:Barnes}) for the same $\Delta_0/\sigma_{\bot}$ and $\omega_I$ as well as for $p_I = 0.4$ and various values of $I$ are shown by dots up to $t = 0.8/\omega_r$. The data in the inset demonstrate the equivalence of the semiclassical approach (Eq.\,\ref{eq:w3}; solid line) and the quantum model (Eq.\,\ref{eq:Barnes} for $p_I = 0.4$ and $I = 3/2$; dots) at early times.}
  \label{fig:Barnes}
\end{figure}

We now turn to the recent non-perturbative solution of the time-convolutionless  master equation \cite{Barnes:2012_PRL} which, within our notation, predicts for an arbitrary shape of the electron envelope function $\psi(\vec{r})$ at given $N_I$ and in the regime $\Delta_0 \gg \sigma_{\bot}$ and $t \ll N_I/|x_IN_0J|$,
\begin{equation}
w_t = \frac{\exp\{it\Delta_0 + [it\Delta_0 +\exp(-it\Delta_0) -1]
y\coth(y/2)/2\Delta_0^2\}}{\cos(\omega_rt) - i\coth(y/2)\sin(\omega_rt)}.
\label{eq:Barnes}
\end{equation}
where $t$ and $\Delta_0$ are in the units of $\sigma_{\bot}^{-1}$ and $\sigma_{\bot}$, respectively;
$y$ is defined in Eq.\,\ref{eq:pI}; $\omega_r = y/2\Delta_0$, and $\Delta_0$ should be replaced by $\Delta_0 -\omega_I$ if $\omega_I \neq 0$. Since, abandoning the relative units, $\omega_r =(x_IN_0J/2N_I)(\Delta_0-g^*\mu_BB)/\Delta_0$, we see that as $t \ll N_I/|x_IN_0J|$, the quantum approach in question is valid for $t \ll |\omega_r^{-1}|$.

As shown in Fig.\,\ref{fig:Barnes}, our theory for arbitrary $p_I$ and $I$ is in a quantitative agreement with the quantum expectations in the regime of their validity.

Finally, we comment on the role played by the field-induced precession of the bath spins. If the corresponding frequency is in the range $1/T_2 \ll  \omega_I \ll \bar{\Delta}$, the adiabatic conditions are fulfilled, so that the precession gives rise to geometrical Berry's phase. Our evaluations with the help of Eq.\,\ref{eq:w3} indicates that in the regime $\Delta_0 \gg \sigma$ the Berry phase contribution to the spin dephasing rate is $\delta\gamma \simeq \omega_I\sigma^2/\Delta_0^2$, where $\gamma$ is twice the inverse of time corresponding to $                    |w_t| = 1/e$. Interestingly, this dephasing mechanism cannot be removed by Hahn's spin echo sequence \cite{Leek:2007_S}.

In summary, we have developed a semiclassical approach to dynamics in the central spin problem for $t < T_2$ and $t < N_I/|x_IN_0J|$, which is in quantitative accord with quantum predictions.  This finding substantiates the applicability of semiclassical models, such as the Landau-Lifshitz-Gilbert equation, to simulations of spintronic nanostructures with a few dozen or more spins. Our physically transparent and mathematically simple model allows  incorporating interactions among bath spins, polaronic effects, and Berry's phase contribution as well as extending the theory to the region of small spin splittings, $\Delta_0 \lesssim \sigma$, so far unaccessible to the quantum approaches. An extension of our theory to longer times, explored already in some quantum models \cite{Coish:2010_PRB,Coish:2008_PRB}, requires consideration of electron-induced bath dynamics and incorporation of corrections to the gaussian approximation, both leading to a dependence of the results not only on $N_I$ but also upon the shape of the electron envelope function $\psi(\vec{r})$.   Another important line of research concerns dynamics of a confined hole interacting with a bath of nuclear \cite{Wang:2012_PRL,Li:2012_PRL,Greilich:2011_NP} or magnetic spins, particularly in view of current interest in dynamics of trapped excitons in DMSs \cite{Klopotowski:2011_PRB,Sellers:2010_PRB,Beaulac:2009_S,Toropov:2006_PRB,Mackowski:2004_PRB,Dorozhkin:2003_PRB,Kavokin:1999_PRB}.

The author thanks {\L}ukasz Cywi\'nski for many valuable discussions and for a critical reading of the manuscript.

%

\end{document}